\documentclass[amsmath,twocolumn,superscriptaddress,prl,aps]{revtex4}
\usepackage{amsthm,amsfonts,graphicx,verbatim, color}
\usepackage{graphicx}
\usepackage{bm}
\usepackage[utf8]{inputenc}
\let\oldmarginpar\marginpar
\renewcommand\marginpar[1]{\-\oldmarginpar[\raggedleft\footnotesize #1]%
{\raggedright\footnotesize #1}}

\newcommand{\be}{\begin{equation}}
\newcommand{\ee}{\end{equation}}
\newcommand{\bea}{\begin{eqnarray}}
\newcommand{\eea}{\end{eqnarray}}

\renewcommand{\epsilon}{\varepsilon}

\renewcommand{\cite}[1]{[\onlinecite{#1}]}

\begin{document}

\title{Localization protected quantum order}
\author{David A. Huse}
 \affiliation{Princeton Center for Theoretical Science, Princeton University, Princeton, New Jersey 08544, USA}
 \affiliation{Department of Physics, Princeton University, Princeton New Jersey 08544, USA}
\author{Rahul Nandkishore}
 \affiliation{Princeton Center for Theoretical Science, Princeton University, Princeton, New Jersey 08544, USA}
\author{Vadim Oganesyan}
 \affiliation{Department of Engineering Science and Physics, College of Staten Island, CUNY, Staten Island, NY 10314}
\affiliation{The Graduate Center, CUNY, 365 5th Ave., New York, NY, 10016}
\author{Arijeet Pal}
 \affiliation{Department of Physics, Harvard University, Cambridge MA 02138, USA}
\author{S. L. Sondhi}
 \affiliation{Department of Physics, Princeton University, Princeton New Jersey 08544, USA}

\begin{abstract}
Closed quantum systems with quenched randomness exhibit many-body localized regimes wherein they do not equilibrate even though
prepared with macroscopic amounts of energy above their ground states. We show that such localized systems can order
in that individual many-body eigenstates can break symmetries or display topological order in the infinite volume limit. Indeed,
isolated localized quantum systems can order {\it even} at energy densities where the corresponding thermally equilibrated system is disordered,
i.e.:  localization protects order. In addition, localized
systems can move between ordered and disordered localized phases via non-thermodynamic
transitions in the properties of the many-body eigenstates. We give evidence that such transitions may proceed via localized
critical points. We note that localization 
provides protection against decoherence that may allow experimental
manipulation of macroscopic quantum states. We also identify a `spectral transition' involving a sharp change in the spectral statistics of the many-body Hamiltonian.
\end{abstract}
\maketitle

Our current understanding of the phases of quantum matter in equilibrium
is built largely on the traditional Landau framework of broken symmetries \cite{LandauLifshitz} and the more recent framework, still in rapid
evolution, of topological order and allied classifications \cite{Wen, Kitaevper, Ludwig, Xie, KitaevSPT, AshvinSPT}. There are
interesting exceptions to these in the presence of quenched
randomness: such as Anderson localization \cite{Anderson}, which is firmly
established in studies of non-interacting particles \cite{localizationRMP}. Recently, the work of Basko, Aleiner
and Altshuler \cite{BAA} and others \cite{gmp,Oganesyan,znid,Warzel,pal} has added to these a previously conjectured \cite{Anderson} extension of Anderson localization to closed, quantum interacting systems---the phenomenon now known as many-body localization (MBL).

To understand the nature of many-body localization, it is useful to first refer to the eigenstate thermalization hypothesis (ETH)  \cite{Deutsch, Srednicki, Rigol}.  The ETH, when
true, applies to the exact many-body eigenstates of the Hamiltonian of a closed, isolated quantum system, in the limit of many degrees of freedom.  The ETH postulates that for
a large class of quantum systems, the probability operator (a.k.a. the reduced density matrix) for any subsystem is, in any exact many-body eigenstate of the full system, equal to the equilibrium Boltzmann-Gibbs distribution at the temperature set by the energy density of the eigenstate.  This occurs because the remainder of the full system
successfully acts as a thermal bath for the subsystem in question.

The many-body localization phase transition is an {\it eigenstate phase transition} from the thermal phase where the exact many-body eigenstates obey the ETH, to the localized phase where the eigenstates violate the ETH;
the latter fail to be a heat bath that thermally equilibrates its subsystems \cite{Anderson, BAA, pal, Oganesyan}. Thus this is a dynamic, but not thermodynamic, phase transition from the thermal phase where the system does thermally equilibrate under the dynamics due to its own Hamiltonian, to the `glassy' localized phase where the isolated quantum system can remain far from thermal equilibrium forever. In a related diagnostic, MBL eigenstates generally display an `area law' entanglement entropy \cite{EE}, unlike thermal eigenstates where the entanglement entropy is generally a `volume law' that reproduces
the thermodynamic entropy.

\begin{figure}
\includegraphics[width = \columnwidth]{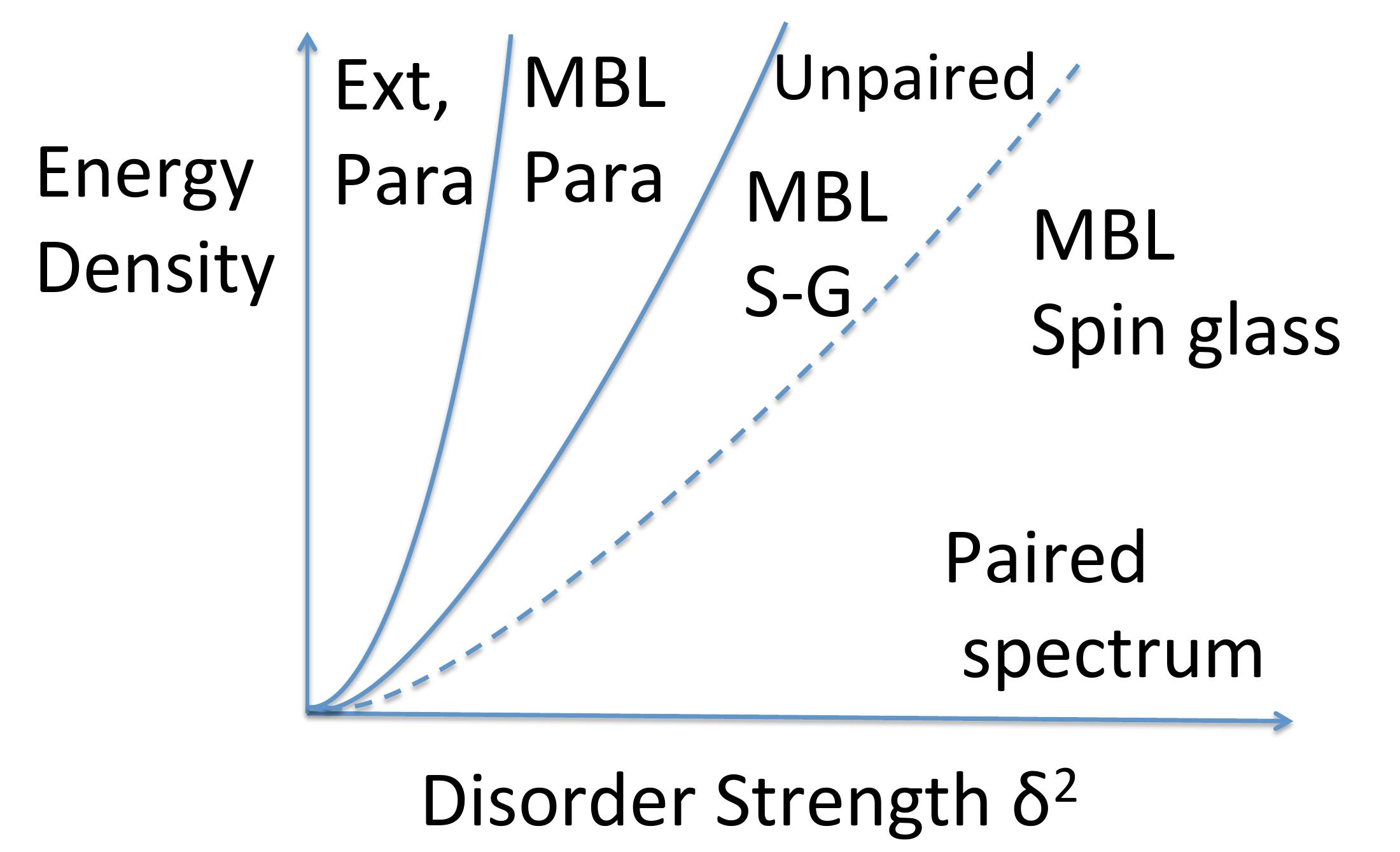}
\caption{\label{fig: 1dphasediagram} In the one-dimensional models we consider, eigenstate properties depend on the energy density, the typical value of $h/J$, the disorder strengths $\delta_J$ and $\delta_h$, and the fermion-fermion interaction strength $\lambda$. This figure shows, schematically, a slice through the phase diagram, for typical $h/J < 1$, and non-zero $\lambda$. Extended (Ext) and many body localized (MBL) phases are separated by the localization transition, ordered spin glass (S-G) and paramagnetic (Para) phases are separated by an eigenstate phase transition, and regions with paired and unpaired spectra are separated by a spectral transition.}
\end{figure}

In this work we examine the highly excited eigenstates, defined as eigenstates with a macroscopic energy above the ground state, of MBL systems and point out that they
come in many flavors, and may be classified in terms of broken symmetries, topological order and/or criticality, very much as in the usual account of phases and
phase transitions in equilibrium systems.
We note that in the presence of many-body localization, equilibrium constraints on order can be evaded: symmetry breaking can occur in highly-excited states of one-dimensional systems, and topological order can arise {\it even in the absence of a bulk gap}.  Instead it is the localization that `protects' the topological order.
We suggest that there can even arise continuous phase transitions between distinct MBL phases, which proceed via MBL critical points. Finally, we point out the existence of yet another kind of transition within the MBL phase: a `spectral transition', which does not involve a change in the properties of the eigenstates, but instead a change in the spectral statistics of the system's Hamiltonian. We emphasize that unlike standard discussions of quantum phase transitions, our discussion is not about ground states or
low-lying excited states, but is about highly-excited eigenstates at energies that would correspond to nonzero (even infinite) temperature if the system could thermalize
at these energies.  This might be useful for experiments, in particular for experiments designed to exploit topological order.

This article is structured as follows:  We first consider three equivalent "integrable" \cite{integrablefootnote} one-dimensional systems - the Majorana chain, $p$-wave superconducting
chain, and transverse-field Ising chain \cite{Fendley}.  The eigenstates of the disordered versions of these Hamiltonians are localized and can be described in terms of noninteracting localized fermions. This localization is robust to weak fermion-fermion interactions \cite{BAA}. The MBL eigenstates can break symmetries, and a robust notion of topological order can be defined for them even without a gap.
We also show how by tuning parameters of the Hamiltonian, we can drive an eigenstate phase transition from one MBL phase to another. In the Ising chain, the transition is of the symmetry-breaking type, from a disordered paramagnet to a phase where the eigenstates are feline (Schrodinger cat) states with long-range spin-glass (SG) order.  In the Majorana and Dirac fermion systems, the transition is between MBL states with and without topological order. We further argue that the critical point separating these two distinct MBL phases can itself be MBL.  We close by discussing extensions of these ideas to higher dimensions. We emphasize again that the ideas discussed here apply only to closed quantum systems - i.e. it is essential that the system not be put in contact with an external thermal bath.

\noindent
{\bf The models:} The three non-interacting one-dimensional models studied in this paper
(to which we will add interactions perturbatively), are the transverse Ising, Majorana, and Dirac fermion chains of finite length $L$ sites, with open ends:
\begin{eqnarray}
H_{\rm{Ising}} &=& -\sum_{i=1}^{L-1} J_i \sigma^z_i \sigma^z_{i+1} -  \sum_{i=1}^{L}  h_i \sigma^x_i ~, \label{eq: HIsing}\\
H_{\rm{Majorana}} &=& -\sum_{j=1}^{L-1} i J_j b_j a_{j+1} -  \sum_{j=1}^{L} i h_j a_j b_j ~, \label{eq: HMajorana}\\
H_{\rm{Dirac}} &=&-\sum_{j=1}^{L-1}  J_j \big( c^{\dag}_j c^{\dag}_{j+1} + c^{\dag}_j c_{j+1} + h.c.\big) \nonumber\\&-&  \sum_{j=1}^{L} h_j (1 - 2 c^{\dag}_j c_j) ~, \label{eq: HDirac}
\end{eqnarray}
where the $\sigma^{x,z}$ are Pauli matrices, 
the $a, b$ are (self-adjoint) Majorana fermion operators and the $c, c^{\dag}$ are conventional Dirac fermion operators. The parameters $J_i$ and $h_i$ are drawn from distributions $P(J)$ and $P(h)$ with means
$\overline J>0$ and $\overline h>0$, and variances $\delta_J^2$ and $\delta_h^2$. The precise details of the distribution are unimportant for our present purposes,
but it is vital that at least one of the variances be non-zero. For specificity, take the distributions to be log-normal, so all $J_i$ and $h_i$ are positive.
We consider the disorder strengths $\delta^2$ to be parameters in our analysis.
These three Hamiltonians are related to one another by the duality transformations \cite{Kogut, Fendley}
\begin{eqnarray}
a_k &=& c_{k} + c^{\dag}_k = \left(\prod_{j<k} \sigma^x_j \right) \sigma^z_k ~, \nonumber\\ \qquad b_k &=&-i (c^{\dag}_k - c_k) =  \left(\prod_{j< k} \sigma^x_j \right)\sigma^y_k ~. \label{eq: Majorana}
\end{eqnarray}
The above Hamiltonians all possess a global $Z_2$ symmetry, implemented by the operator $\hat P$, which takes the form
\begin{equation}
\hat P = \prod_{j=1}^L \sigma^x_j =  \prod_{j=1}^L i a_j b_j =  \prod_{j=1}^L (1 - 2 c^{\dag}_j c_j) ~.
\end{equation}

In the ordered phase (typical $h/J <1$), there is an `edge mode' operator $\hat O^{\dagger}$, whose commutator with the Hamiltonian is
exponentially small ($(h/J)^L\ll 1$) in $L$.  In the Majorana language \cite{Fendley},
\begin{eqnarray}
\hat O^{\dagger} &=& a_1+ i b_L + \frac{h_1}{J_1} a_2 + i \frac{h_{L-1}}{J_{L-1}} b_{L-1} + \frac{h_1h_2 }{J_1J_2} a_3 \nonumber \\
&+& i \frac{h_{L-1}h_{L-2}}{J_{L-1}J_{L-2}} b_{L-2} + ... \label{eq: zero mode}
\end{eqnarray}
This operator creates a Dirac fermion which is {\it bilocalized} around the two edges, and has an energy which is exponentially small in the system size, $E \sim \exp \big(-L \ln(J/h) \big)$.  The existence of this quasi-zero-energy edge mode can be considered a diagnostic for topological order. Since $\hat O$ and $\hat P$ anti-commute, acting on a state with $\hat O$ or $\hat O^{\dagger}$ flips its eigenvalue under $\hat P$.

\noindent
{\bf Localization in the ordered phase:} To discuss localization in the non-interacting model, it is convenient to use the Ising spin formulation (\ref{eq: HIsing}). The results apply equally to the Majorana and Dirac fermion chains.

In the ordered phase, $\overline J - \sqrt{\delta^2} \gg \overline h$, the ground state is a ferromagnet, with long-range order of $\sigma^z$.  It is even under $\hat P$, consisting of a linear combination of states with the average $z$-magnetization `up' and `down'.  Let's call it $|0,+\rangle$.  It is a `feline' (Schrodinger cat) state consisting of a coherent linear combination of two `macroscopically' different states.  If we let $\hat O$ operate on this ground state, this makes the other `ground state', $|0,-\rangle=\hat O|0,+\rangle$, which is also feline, is odd under $\hat P$, and is higher in energy by an amount that is exponentially small in $L$.

The normal modes in the ordered phase are fermion operators, which anticommute with each other and with $\hat P$. The action of these normal modes on the spin state is straightforward: they create domain walls (with respect to the ferromagnetic ground state). Domain walls sit on bonds, and are created by the self-adjoint operators
\begin{equation}
d_{k+1/2} = \left(\prod_{j<k} \sigma^x_j\right) = \left(\prod_{j<k} i a_j b_j\right) = \left(\prod_{j<k} (1 - 2 c^{\dag}_j c_j)\right) \label{eq: domain walls}
\end{equation}

In the integrable Hamiltonian, the properties of an eigenstate may be extracted from the behavior of the normal modes, which create domain walls. In the classical limit $h = 0$, the normal modes are trivially localized. Away from the classical limit, they hop under the action of the transverse fields $h_i$, and see a spatial potential $J_i$. In the clean limit, the normal modes (and hence the domain walls) are delocalized over the chain. However, it is well known that non-interacting fermions hopping in a one-dimensional random potential always experience localization \cite{localizationRMP}. Therefore, in the presence of non-vanishing disorder in $J_i$, the normal modes (and hence the domain walls) must necessarily be localized. In the strong disorder limit, when $\delta_J^2 \gg \overline h^2$, the single domain wall eigenstates 
are exponentially localized, with localization length
\begin{equation}
\xi_{loc} \sim \frac{1}{ \ln |\delta_J^2/\overline h^2|}
 \label{eq: length1}
\end{equation}
This follows straightforwardly from perturbation theory in small $h/\delta_J$.
Localization of the domain walls in each many-body eigenstate implies that each eigenstate has long-range spin glass order \cite{SG}, i.e. the correlation function $\langle \sigma^z_0 \sigma^z_r\rangle$ within the eigenstate is non-zero for large $r$, with its sign set by how many domain walls are present and localized between sites $0$ and $r$.
Observe that this spin-glass order breaks the global $Z_2$ symmetry in these highly-excited localized eigenstates of this one-dimensional system,
although at thermal equilibrium such discrete symmetry breaking at non-zero temperatures is forbidden. This eigenstate spin-glass ordering is also discussed in Pekker, {\it et al.} \cite{Pekker}.

Just as this model in this ordered phase has two nearly-degenerate ground states $|0,\pm\rangle$ that are each feline, the excited eigenstates also come
in nearly-degenerate feline pairs $|n,\pm\rangle$, produced by adding some particular set of localized domain walls, $n$, to each of the two ground states.
The states $|n, \pm\rangle$ are orthogonal eigenstates of the global spin flip operator $\hat P$, with eigenvalue $\pm1$. From these two eigenstates we can make the state $|n,\uparrow\rangle\equiv(|n,+\rangle+|n,-\rangle)/\sqrt{2}$ which has some particular pattern of local $z$-magnetizations dictated by the localized domain walls. Each eigenstate $|n,\pm\rangle$ is a (feline) coherent linear combination of this spin-glass state $|n,\uparrow\rangle$ with the opposite state under the global spin flip operation, $|n,\downarrow\rangle=\hat P|n,\uparrow\rangle$. The energy splitting between $|n, \pm \rangle$ is exponentially small in system size, and thus the energy uncertainty in the symmetry-breaking states $|n,\uparrow \rangle$ and $|n, \downarrow \rangle$ vanishes exponentially with the number of spins in the infinite volume limit. We note in passing that the existence of spin glass order in the eigenstates is consistent with exponential localization of the local fluctuations of conserved quantities - a hallmark of the MBL phase.

\noindent
{\bf Robustness of localization to weak interactions:}
 We assume we are working with highly-excited states, where the localized domain walls are dense and
 frequently overlap (the typical domain wall separation can be less than a
 localization length) \cite{footnote2}.  We then add to the Hamiltonian weak short-range interactions between these domain walls,
 and ask whether they cause a breakdown of localization.
 This situation is essentially identical to that analyzed by Basko, {\it et al.} \cite{BAA}, who show that many-body
 localization is stable to weak nonzero interaction $\lambda$ as long as the localization length of the single-particle states is finite.  We also assume that the interactions commute
with $\hat P$ and thus respect the $Z_2$ symmetry.

If the disorder is weak, the localization will be destroyed by strong enough interaction,
making the extended, thermal phase at weak disorder and high energy density, as shown in Fig. 1.  In terms of the Ising model, in the extended phase
the spin-spin correlations within a thermal eigenstate are short-range.  If we assume the eigenstate phase transitions are continuous (not first-order),
then there must be two transitions, as in Fig.1.  Starting in the extended thermal phase and increasing the disorder, first we reach the
localization transition.  Once we enter the localized phase, the spin-spin correlations in the eigenstates can start increasing to longer range
than they are at thermal equilibrium.  The transition to the MBL spin-glass phase is when the spin correlations develop long-range order.

\noindent
{\bf Spectral transition in the ordered phase}:
In the MBL spin-glass phase, all eigenstates come in parity-related pairs $|n,\pm\rangle$ which differ only by their occupation of the single
particle edge mode created by $O^{\dag}$.  The energy of the edge mode is exponentially small in the chain length, $E_{O} \sim \exp(-L/\xi)$, where
$\xi$ is the localization length of the edge mode. Thus, the level splitting within such a pair is $E_{O} \sim \exp(-L/\xi)$. Meanwhile, the typical
many-body level spacing is of order $\delta E \sim \exp(- sL)$, where $s$ is the thermodynamic entropy per site that would result if the system
equilibrated at the given energy density.  Thus, tuning $\xi$ or $s$ leads to a spectral transition.  For strong disorder and thus small localization
length $\xi$ and/or for low energy density and thus low entropy $s$, the spectrum is `paired'.  Here the edge mode level splitting is less than the typical
level spacing so the many-body spectrum at large $L$ consists of nearly-degenerate doublets with Poisson inter-doublet level spacing statistics. 
The level spacing within each doublet is exponentially small in system size compared to the typical level spacing between doublets. Meanwhile, for
weaker disorder and/or higher entropy, the spectrum is unpaired. In this unpaired regime, the energy of the single particle
edge mode is exponentially larger than the
typical many body level spacing, and as a result we have Poisson statistics for all of the individual many-body energy levels.  The Poisson (or paired Poisson)
level statistics themselves are a diagnostic for MBL, since in the thermal phase, the level statistics are instead characteristic of the Gaussian orthogonal
ensemble (GOE). 

\begin{figure}
 \includegraphics[width = \columnwidth]{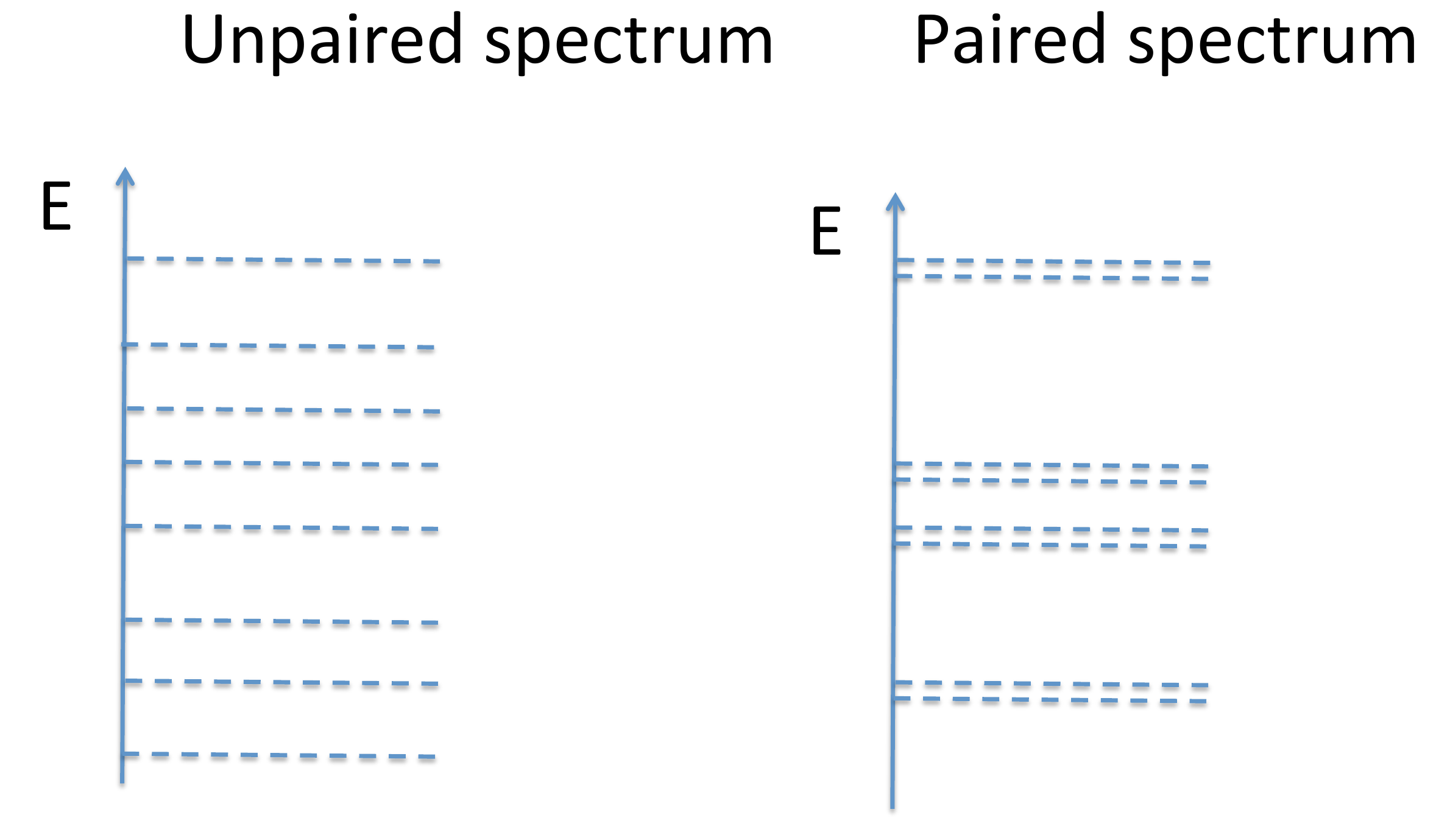}
 \caption{\label{fig: Spectrum} An illustration of the spectral transition between paired and unpaired many-body spectra, driven for example by tuning disorder strength.}
 \end{figure}

The spectral transition is illustrated in Fig. \ref{fig: Spectrum}. It does not involve a change in the symmetry or topological order of the eigenstates, but rather involves a change in the many-body spectral statistics of the Hamiltonian. It may be detected in numerics by examining the ratio of two consecutive gaps, $r = \mathrm{min}(\delta_n, \delta_{n+1})/\mathrm{max}(\delta_n, \delta_{n+1})$, where $\delta_n$ is the energy gap between the $n^{th}$ and $(n+1)^{th}$ many body eigenstates. This ratio will have average value $(2 \ln 2 - 1)$ in the `unpaired' localized regime \cite{Oganesyan}, whereas the average value in the paired regime will be exponentially small in system size (zero in the infinite volume limit). Thus the ratio of two consecutive gaps provides a sharp diagnostic for the spectral transition.

\noindent
{\bf Magnetic response}:
  One curious feature of these pairs $|n,\pm\rangle$ of feline eigenstates is that the states $|n,\uparrow\rangle$ have a typical $z$-magnetization $m$ that scales as $\sim\sqrt{L}$.  A longitudinal magnetic field $B$ added to the Hamiltonian thus acts on the Hilbert space of a pair (in the $B=0$ eigenstate basis) as
 \begin{equation}
 H_{eff} \sim \left( \begin{array}{cc} E & B \sqrt{L} \\ B\sqrt{L} & E+\bar h e^{-L/\xi} \end{array} \right)
 \end{equation}
It is easy to verify that the energy eigenstates have an adiabatic magnetic susceptibility $\chi = \left(\partial m/\partial B\right)_{B=0}$ that is exponentially
large in $L$, but is of opposite sign between the two paired eigenstates.
The feline nature of the eigenstates is thus destroyed by an exponentially small (in $L$) longitudinal field.
Interestingly, this makes this system a possibly very sensitive magnetic field sensor.  If a state with a nonzero $z$-magnetization is prepared,
it will be localized and stable as long as the magnitude of the longitudinal field $B$ is larger than $\sim e^{-L/\xi}$.
However, once the magnitude of the longitudinal field drops below this level, then the system's many-body eigenstates, although still localized,
are feline and the magnetization will decrease.
The main caveat to this apparently exponentially-fine-in-$L$ field sensitivity is that the resulting magnetization dynamics is exponentially slow in $L$, and the utility of this idea will be limited by the decoherence rate of any real system. We note however that, in our three dimensional world, three dimensional MBL systems where only the boundary is coupled to the environment should be protected from environmental decoherence, since the effects of the environment should only propagate into the system up to one localization length.  

\noindent
{\bf Localization in the paramagnetic phase}: We now turn our attention to the disordered regime, $\overline h - \sqrt{\sigma_h^2} \gg \overline J$.
In this regime, the (Ising) ground state has all spins aligned on average along the $x$ axis, and the elementary excitations are spin flips.
A similar analysis to above leads one to conclude that all spin flips are localized in the non-interacting model, as long as $\delta_h^2 \neq 0$.
Localization is robust against weak enough interactions so long as the localization length $\xi$ is finite. 
Thus, the eigenstates of the disordered Ising paramagnet can also be MBL.  However, these eigenstates do not break $Z_2$ symmetry and do not come in pairs. Likewise, there is no edge mode in the fermion language, and no topological order.

These results can also be understood in terms of the well known self-duality of the one dimensional Ising model \cite{Kogut}, which swaps $h$ and $J$. In particular, it follows from self duality that if the ordered phase is MBL, the disordered phase is also MBL. However, the spectrum in the paramagnetic phase is not paired. This is because the pair of parity related MBL eigenstates $|n,\pm \rangle$ on the ordered side differ by the edge mode $\hat O$, and the self duality acts non-locally on the edge mode, mapping the `parity related pair' of ordered states to disordered states with different boundary conditions. Thus, the disordered side (with specified boundary conditions) has states that do not come in pairs and do not have edge modes, and are separated from their counterparts on the ordered side by a phase transition.

\noindent
{\bf The phase transition:} In the absence of interactions ($\lambda = 0$) there is a continuous transition between localized phases, which proceeds via an infinite randomness critical point {\it even} for highly excited states. We begin by discussing this non-interacting picture, before examining the effect of interactions. The critical regime is treated using the strong disorder renormalization group (SDRG) \cite{Fisher}. The SDRG proceeds by sequentially identifying the strongest bond or field in the Hamiltonian, diagonalizing it, and determining the coupling to the rest of the system perturbatively.  This procedure can be shown to be asymptotically exact, because the system flows to strong disorder.

When looking for ground states \cite{Fisher}, after diagonalizing a particular bond or field, we should truncate to the lower energy subspace. However, this method can be easily generalized to obtain excited states, by sometimes truncating into the higher energy subspace instead. It can be verified that this does not change the flow equations for $|J_i|$ or $|h_i|$, but merely introduces some extra minus signs (some bonds become antiferromagnetic and some fields point along $-x$). We thus recover the flow equations \cite{Fisher} for the probability distributions $P(h)$ and $P(J)$ as flow equations for $P(|h|)$ and $P(|J|)$. The flow \cite{Fisher} is to strong disorder, so the transition should proceed via an infinite randomness critical point, for all excited states.

In either phase we have already argued that the non-interacting system is localized. However, at criticality ($\epsilon = 0$), the spectrum of the non-interacting Hamiltonian contains states with all localization lengths and the single-fermion states in the limit of zero energy have a diverging localization length. Nonetheless, the critical point of the non-interacting Hamiltonian (\ref{eq: HIsing},\ref{eq: HMajorana},\ref{eq: HDirac}) is localized in the following sense. Almost all single-fermion wave functions are localized, with only the limiting zero-energy states having infinite localization length. The entanglement entropy (within a many-body eigenstate) of a subregion is thus sub-extensive, unlike a thermal state, which has extensive entanglement entropy \cite{EE}. In fact, since the RG flow for $P(|J|), P(|h|)$ is the same in the excited eigenstates as for the ground state, the entanglement entropy is also the same in the excited eigenstates as in the ground state. As determined in \cite{Refael}, the entanglement entropy of a subregion at the infinite randomness critical point has a leading term that scales as $\ln L$, whereas a thermal state would have an entanglement entropy that scales as $L$. Thus, the (non-interacting) critical point violates the ETH, and can be considered localized.

We now discuss whether localization can survive interactions at the critical point. The key question is whether the presence of a subextensive number of critical modes can cause the rest of the degrees of freedom to delocalize, by mediating resonances between distant near-degenerate localized modes. In \cite{Vosk}, it was determined that the mediated interactions fall off exponentially with distance, whereas the typical level spacing decreases as a power law of distance. Thus, the mediated interaction between distant near-degenerate modes should be weaker than the level splitting, and mediated interactions should be unable to delocalize the formerly localized degrees of freedom. This argument suggests that the critical point separating two MBL phases can itself be MBL. A slice through the phase diagram is presented in Fig.\ref{fig: 1dphasediagram}. For more detail on the phase transition within the MBL regime, see \cite{Pekker}.

\noindent
{\bf Topological order in the fermion language:} Translating to the Majorana and Dirac fermion languages, we again conclude that all bulk modes are localized, with localization lengths equal to those calculated above. The bulk normal modes are Dirac fermions. Meanwhile, the nearly-degenerate pairs of excited eigenstates are states where the single particle edge mode created by $\hat O^{\dagger}$ is either
occupied or unoccupied.  Thus, the eigenstates of (\ref{eq: HMajorana},\ref{eq: HDirac}) have topological order.
The existence of topological order follows trivially in these non-interacting models, because the edge mode does not interact with anything in the bulk, and is bilocalized as two Majorana modes, one near each end of the chain. In the clean system, the addition of arbitrarily weak interactions destroys the topological order, since the Majorana end modes can couple through the delocalized domain walls. However, in the disordered system, we will show that many body localization protects the topological order.

\noindent
{\bf Localization protects topological order:} We now demonstrate that the edge Majoranas remain localized in the presence of interactions, even in the absence of a gap. We assume that the interaction locally conserves fermion number modulo 2, and is thus a product of an even number of Majorana operators. Before turning on interactions, there are an odd number of quasi-zero-energy Majorana modes localized at each edge. We cannot turn this into an even number of Majoranas by acting with an even number of Majorana operators, so the edge Majorana must survive the addition of interactions. The Majorana cannot disappear because of hybridization with localized bulk modes because the bulk modes are Dirac. The only way to make the topological order disappear is to couple the two Majoranas at either edge. However, the interaction cannot do this because it is short range, and the Majorana cannot be passed from one bulk Dirac mode to another, until it reaches the other Majorana at the opposite end of the chain, {\it because the bulk is MBL}, and does not allow energy or particles to propagate. Thus, MBL in the bulk protects topological order in highly-excited localized states, just as a bulk gap can protect topological order in ground states. This is in sharp contrast to the clean (non-MBL) system, where the Majorana edge modes can hybridize with each other through delocalized bulk modes in any excited state. The localization protection of the edge Majoranas in quantum states other than ground states might be useful for experiments designed to exploit topological order. In particular, it might be useful for experiments designed to detect Majorana fermions in quantum wires \cite{Franz}.

\begin{figure}
\includegraphics[width = \columnwidth]{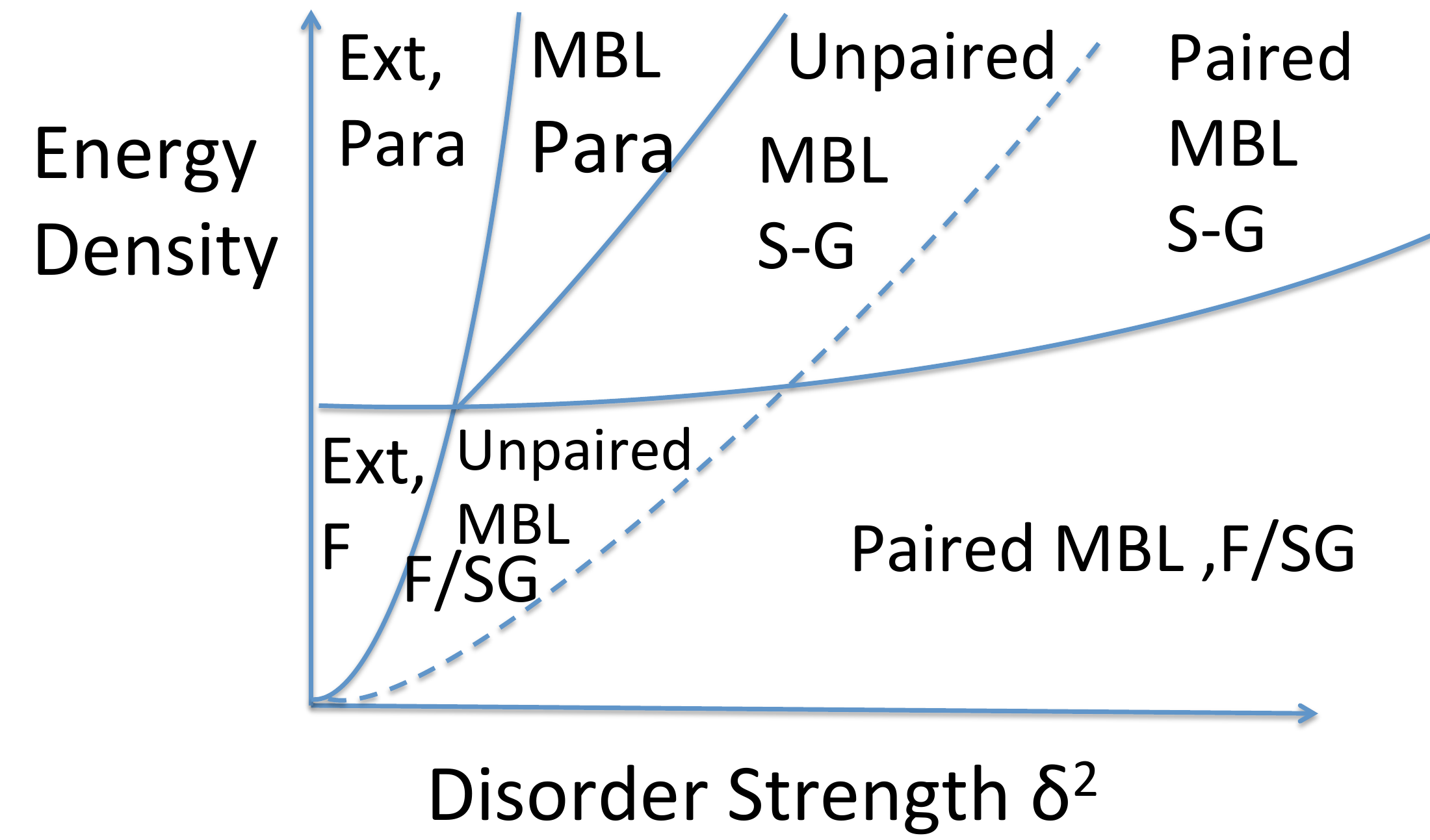}
\caption{\label{fig: 2dphasediagram} A (schematic) slice through the $(d>1)$-dimensional phase diagram at non-zero $\lambda$ and typical $h/J <1$. In addition to the eigenstate phase transitions and spectral transition present in one dimension (Fig. \ref{fig: 1dphasediagram}), now there is also the usual thermodynamic phase transition between phases with and without ferromagnetic order, i.e. net spontaneous magnetization. The presence of ferromagnetic order is labelled by $F$.}
\end{figure}

\noindent
{\bf Symmetry Breaking in $d \ge 2$:} The ideas discussed above have a straightforward extension to the Ising model in more than one dimension. A major difference is that in two or more dimensions,
thermodynamic $Z_2$ symmetry breaking persists to nonzero excitation-energy densities even in the extended (thermal) phase.  As a result, the phase diagram contains another type of
phase transition - the usual thermodynamic phase transition between states with and without ferromagnetic order. A slice through the $(d>1)$-dimensional phase
diagram is presented in Fig. \ref{fig: 2dphasediagram}.

As in Fig. 1, we assume that the eigenstate phase transitions are continuous (not first-order).
This again implies that a localized paramagnetic phase exists between the extended (thermal) paramagnetic phase and the localized spin glass in the
higher-energy regime above the ferromagnetic phases.  Moving across this phase by increasing disorder, the spin-spin correlation length grows continuously from the
finite thermal value of the extended phase and diverges at the transition to the spin glass.
We note that this symmetry-breaking paramagnet-to-spin-glass phase transition in the localized Ising model in $d \ge 2$ is also governed by an infinite randomness
fixed point \cite{motrunich,others} which again should extend to finite energy density states and lead to sub-thermal entanglement.
We believe that localization persists here too, although a more detailed analysis is desirable.  Assuming the symmetry-breaking transition in the localized
phase out of the paramagnet is indeed in the infinite-randomness universality class, there should be localized domain walls in the ordered phase just across the
transition.  The higher-energy eigenstates with these domain walls present are spin-glass states.  This is why we believe the spin-glass phase always exists between the paramagnet and the ferromagnet in the localized regime, as shown in Fig. 3.
We also note that unlike in
$d=1$, the nearest-neighbor transverse-field Ising model in $d \ge 2$ {\it is} an interacting system on its own.

\noindent
{\bf Topological Order in $d \ge 2$:} From our results thus far we can immediately draw some interesting conclusions about
topologically ordered systems in $d \ge 2$ by simply dualizing the Ising model. This leads to MBL protected topological order in higher dimensions, as we now discuss.

Let us begin in $d=2$ where the ``Ising model'' of a system that exhibits topological
order is the the $Z_2$ lattice gauge theory with matter,
governed by the Hamiltonian \cite{Kogut, Fradkin, FM}
\begin{eqnarray}
\label{IGTHamil}
-H = &&\sum_p K_p \prod_{l \in \partial p} \sigma^z_l +
 \sum_l \Gamma_l \sigma^x_l \\ \nonumber
&+&  \sum_l J_l \sigma^z_l \prod_{s \in \partial l} \tau^z_s
+  \sum_s \Gamma^M_s \tau^x_s
\end{eqnarray}
supplemented by the constraint that we restrict its action
to ``gauge invariant'' states defined by
\begin{equation}
G_s |\psi \rangle = | \psi \rangle \ \ \ , \ \ G_s= \tau^x_s
\prod_{l: s \in \partial l} \sigma^x_l \ .
\label{eq:IGTconstraint}
\end{equation}
In the above the gauge ($\sigma^i_l$) and  matter ($\tau^i_s$) operators act in spin-$1/2$ Hilbert spaces that
live on the links $l$ and sites $s$ of the square lattice with plaquettes $p$ and $\partial p$ and $\partial l$
are the boundaries of the corresponding objects.
On dualizing the Ising model, we get the $Z_2$ gauge theory {\it without} matter, obtained from (\ref{IGTHamil}) by
dropping the matter degrees of freedom entirely. In this case the parameters $K_p$ and $\Gamma_l$ are, numerically,
the on-site fields and bond interactions of the dual Ising model.

Let us briefly review some salient facts about the non-random system. In this, the paramagnetic and ferromagnetic
phases dualize, respectively, to the (topologically ordered) deconfined and (non-topological) confined phases of the gauge theory
where the terminology refers to the energy needed to separate two test charges to infinity. For our purposes, it is more
useful to consider the
standard equal-time diagnostic which can be evaluated in individual eigenstates, the Wilson loop \cite{Kogut} for a contour C
\begin{equation}
W[C] =  \langle \prod_{l \in C} \sigma^z_l \rangle
\end{equation}
which exhibits a perimeter/area (P/A) law decay, $\log{W[C]} \propto -P/-A$, in the topological/non-topological phase. A useful
picture of the ground states and excitations is obtained by thinking in the basis of eigenstates of $\sigma^x_l$. At
$\Gamma \gg K$, the non-topological ground state has $\sigma^x_l = 1$ on all bonds. The elementary excitations are small loops of bonds where $\sigma^x_l = -1$, which we refer to as bonds with $Z_2$ ``electric'' flux. These loops are dual to domain walls above the ferromagnetic state in the Ising language. As we pass to the topological phase at $\Gamma \ll K$, the loops/domain walls
proliferate and their condensation signals the transition. The elementary excitations in the topological phase are visons,
plaquettes where $\prod_P \sigma^z_l = -1$.

Now the translation is straightforward and we conclude that with sufficient randomness in the couplings there exist both
MBL localized topological and non-topological phases in the $Z_2$ gauge theory without matter.  [In Figure 3, we can simply relabel the
paramagnetic and spin glass phases as topological and non-topological respectively.]  Of maximum interest is the
MBL topological phase whose topological order is protected by localization and would not exist in its absence at
nonzero energy densities (the dual of the thermal paramagnet phase in Fig.3 is a phase where topological order is destroyed by thermal fluctuations). In the MBL topological phase, which is usefully described as a state with a finite density of localized
visons, the eigenstates display a ``spin glass'' version of the perimeter law in which the magnitude of $W[C]$ decays exponentially with perimeter but with a sign that depends on how many localized visons are encircled by the loop. By contrast, in the non-topological phase, $W[C]$ exhibits an area law.

\begin{figure}
\includegraphics[width = 0.75\columnwidth]{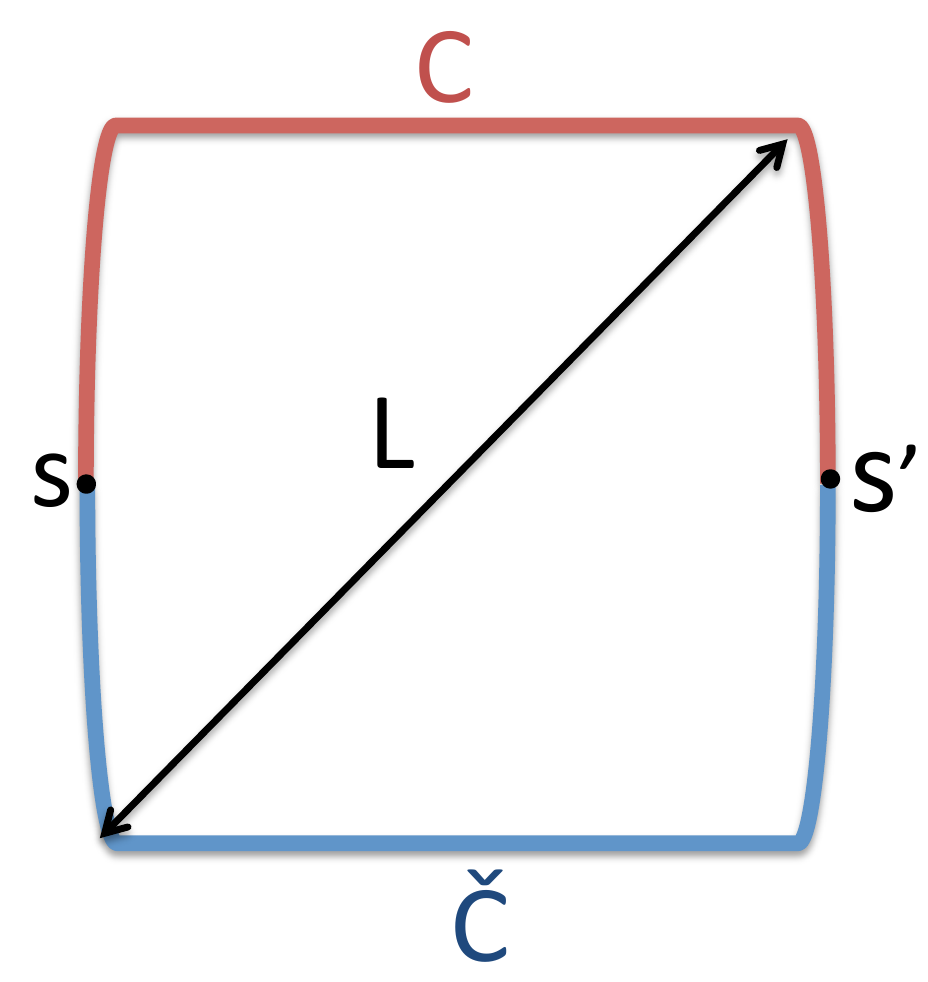}
\caption{\label{fig:FMFigure} The contours used to define the Fredenhagen-Marcu order parameter for a translationally non-invariant system.}
\end{figure}

This account of the $Z_2$ topological phase can be extended in two
directions. First, one can include gauge charged matter, as in (\ref{IGTHamil}),
which is known to leave the clean system topologically ordered at $T=0$
when $J \ll \Gamma^M$. In the presence of sufficient randomness, the matter
excitations will be localized and we will obtain an MBL protected topological
phase in the presence of dynamical matter at nonzero energy densities. However
this phase can no longer be diagnosed by examining the Wilson loop, which
exhibits a perimeter law in all phases in the presence of matter. Instead
we turn to a version of the Fredenhagen-Marcu order parameter \cite{FM}, which measures the ``line tension'' of $Z_2$ electric fields specified by their
distribution in an eigenstate. The version needed for our random setting
is
\begin{equation}
R(L) = \frac{\langle \tau^z_s \big(\prod_{l \in C}\sigma^z_l ) \tau^z_{s'} \rangle
\langle \tau^z_s \big(\prod_{l \in \check{C}}\sigma^z_l ) \tau^z_{s'} \rangle} {\langle \prod_{l \in C \cup \check{C}} \sigma^z_l \rangle}
\end{equation}
where the notation is explained in Figure \ref{fig:FMFigure} and the expectation values are
taken in a particular eigenstate of the Hamiltonian. Extending the discussion
in \cite{FM} to the present case we propose that $R(L\rightarrow \infty) = 0$ in the topological phase, but not in the confined phase.

The second extension
we propose involves two other common diagnostics of topological order in the
clean system---a ground state topological degeneracy of $4^g$ on closed manifolds of genus $g$ and a topological entanglement entropy of $\log 2$ \cite{anushya}.
In the MBL $Z_2$ topologically ordered phase we expect that the dominant
finite size effect in large systems of linear dimension $L$ will arise from
$O(e^{-L})$ tunneling between clusters of $4^g$ finite energy density states
which (roughly) exhibit the same pattern of localized excitations but differ
in the presence or absence of visons threading the non-contractible loops on
the manifold. Likewise individual eigenstates will exhibit an
area law in entanglement and a topological constant piece of size $\log 2$
which can be detected by means of the subtraction procedure outlined, e.g.
in \cite{levinwentent, kitaevpreskill}.  

Dualizing the Ising model in $d > 2$ will yield a d-form
gauge theory \cite{Wegner} with a topological phase with pointlike excitations and
MBL stabilized topological order.  Getting to a more conventional gauge
theory in $d>2$ however requires some fresh thinking. For example, in $d=3$
we need to consider the localization of stringlike excitations (vison loops)
and one cannot simply appeal to the body of results to date in this paper
or elsewhere. 

Finally, we would draw the reader's attention to \cite{Wootton, Stark} where
the dynamical localization of excitations in $Z_2$ gauge theory with matter has 
been discussed in the language of the perturbed toric code relevant to quantum 
information storage.

\noindent
{\bf Conclusions and outlook:} Thus we have shown that eigenstates of MBL systems come in many flavors, and may be classified in terms of broken symmetries, topological order, and criticality, just like extended states. Localization itself can protect order through the intuitive mechanism of localizing excitations that would disrupt it.
We have also identified a new kind of transition - the spectral transition, involving a change in the spectral statistics of the Hamiltonian.
The protection of order and quantum coherence by localization might open the door to a new generation of quantum devices that are immune to environmental noise,
and are not restricted to ground states or low-energy states. It may also be useful for ongoing experiments attempting to observe Majorana fermions in quantum wires \cite{Franz}. 

We have focused, for pedagogical clarity, on broken $Z_2$ global symmetry and $Z_2$ topological order.  Generalizations should be straightforward to other problems
where the elementary excitations subversive of ordering can be localized at all energies by sufficiently random couplings. Immediate examples are $p \ge 3$ $Z_p$ 
clock models in $d \ge 1$ and dual parafermionic systems in $d=1$ \cite{Fendley} and $Z_p$ gauge theories in $d \ge 2$. Farther afield we should flag 
other models with broken discrete symmetries and also topologically ordered systems with discrete gapped excitations such the Levin-Wen models that
exhibit non-abelian phases \cite{LevinWen}. However, it is essential for our purposes that the system should not support long-range interactions, including those which might be mediated by Goldstone modes or gapless gauge excitations (``photons'') that do not localize \cite{Chalker}. Thus, an extension of these ideas to systems with broken 
continuous symmetries and
continuous gauge groups looks problematic. However, an extension to continuous symmetries might be possible if the Goldstone mode were gapped out by the Anderson-Higgs mechanism, or by placing the system on a Bethe lattice, where Goldstone bosons are absent \cite{Laumann}. We defer further consideration of these issues to future work.

While this work has focused on MBL eigenstates, an experimental construction of a system with a Hamiltonian that displays many body localization will necessarily start with an initial state that is not an exact eigenstate of the Hamiltonian. We expect, based on analogies to the `diagonal ensemble' viewpoint advocated in \cite{Polkovnikov}, that for many initial states the density matrix at long times can be treated as being diagonal in the basis of energy eigenstates. However, a detailed understanding of the dynamical evolution of an initial superposition state (or mixed state) is an important topic for future work. 

A final set of interesting open questions involves whether continuous phase transitions between distinct MBL phases necessarily proceed via a localized critical point. We have argued that for MBL phases in the one-dimensional random Ising model, phase transitions between MBL states proceed via an infinite-disorder critical point that is itself MBL. It remains to be determined to what extent this holds true for more complex models. Another open question involves the nature of the phase transition between extended (thermal) and MBL phases, which to our knowledge has not yet been determined in any system.

\noindent
{\bf Acknowledgements:} We would like to thank Ehud Altman, Ignacio Cirac, Arun Nanduri, Anushya Chandran and Gil Refael for discussions and suggestions.  This research was supported in part by the National Science Foundation under DMR 08-19860 (DAH), DMR-0955714(VO) and DMR 10-06608 and PHY-1005429 (SLS), and by the DARPA OLE program (DAH).


\begin{thebibliography}{99}
\bibitem[integrablefootnore (2013)]{integrablefootnote}
We call these many-body systems that possess a representation in terms of non-interacting particles "integrable", although the localized single-particle states that are the "constants of motion" may need to be obtained numerically for any specific realization of the quenched randomness.
\bibitem[footnote2 (2013)]{footnote2}
In the opposite, dilute limit, domain walls do not overlap, do not interact, and localization is trivially robust.
\end{thebibliography}
\end{document}